\documentclass[twocolumn,amssymb, nobibnotes, aps, amsmath, amssymb, superscriptaddress,floatfix]{revtex4-1}

\usepackage[utf8]{inputenc}
\usepackage[T1]{fontenc}
\usepackage{graphicx}
\usepackage{dcolumn}
\usepackage{bm}
\usepackage{newtxtext, newtxmath}  
\DeclareGraphicsExtensions{.pdf,.jpeg,.jpg, .png, .eps, .tiff}
\usepackage{epstopdf}
\usepackage{lipsum}
\usepackage{tabularx}
\usepackage{booktabs}
\usepackage{array}
\usepackage{natbib}
\usepackage{xcolor}
\usepackage{grffile}
\usepackage{boxhandler}
\usepackage{float}
\usepackage[normalem]{ulem}
\usepackage{hyperref}

\renewcommand{\vec}[1]{\boldsymbol{#1}}
\newcommand{\tens}[1]{\boldsymbol{#1}}

\definecolor{mygreen}{rgb}{0.0,0.55,0.3}

\definecolor{strawberry}{rgb}{1.0,0.0,0.5}

\begin{document}
\title{Necking and failure of a colloidal gel arm: signatures of yielding on different length scales}
\author{Kristian Thijssen}
\affiliation{Niels Bohr Institute, University of Copenhagen, Blegdamsvej 17, Copenhagen 2100, Denmark}
\affiliation{Yusuf Hamied Department of Chemistry, University of Cambridge, Lensfield Road, Cambridge CB2 1EW, UK}

\author{Tanniemola B. Liverpool}
\affiliation{School of Mathematics, University of Bristol, Fry Building, Bristol BS8 1UG, UK }

\author{C. Patrick Royall}
\affiliation{H.H. Wills Physics Laboratory, Tyndall Avenue, Bristol, BS8 1TL, UK}
\affiliation{School of Chemistry, University of Bristol, Cantock's Close, Bristol, BS8 1TS, UK}
\affiliation{Gulliver UMR CNRS 7083, ESPCI Paris, Universit\'{e} PSL, 75005 Paris, France}

\author{Robert L. Jack}
\affiliation{Yusuf Hamied Department of Chemistry, University of Cambridge, Lensfield Road, Cambridge CB2 1EW, UK}
\affiliation{DAMTP, Centre for Mathematical Sciences, University of Cambridge,
Wilberforce Road, Cambridge CB3 0WA, UK}

\begin{abstract} 
Colloidal gels consist of percolating networks of interconnected arms.  Their mechanical properties depend on the individual arms, and on their collective behaviour.  We use numerical simulations to pull on a single arm, built from a model colloidal gel-former with short-ranged attractive interactions.  Under elongation, the arm breaks by a necking instability.  We analyse this behaviour at three different length scales: a rheological continuum model of the whole arm; a microscopic analysis of the particle structure and dynamics; and the local stress tensor.  Combining these different measurements gives a coherent picture of the necking and failure: the neck is characterised by plastic flow that occurs for stresses close to the arm's yield stress.  The arm has an amorphous local structure and has large residual stresses from its initialisation.  We find that neck formation is associated with increased plastic flow, a reduction in the stability of the local structure, and a reduction in the residual stresses; this indicates that {the} system loses its solid character and starts to behave more like a viscous fluid. We discuss the implications of these results for the modelling of gel dynamics.
\end{abstract}

\maketitle

\newcommand{\dampbare}{\lambda_0}
\newcommand{\damp}{\lambda}
\newcommand{\sigY}{\sigma_{\rm Y}} 
\newcommand{\taub}{\tau_{\rm b}} 
\newcommand{\sigbarz}{\overline{\sigma}^{zz}}

\section*{Introduction}\label{intro}

Colloidal gels encompass a range of materials, with  diverse applications including food products \cite{mezzenga2005understanding}, tissue engineering~\cite{diba2017highly} and printing technology~\cite{xiong2019}.  
They typically consist of heterogeneous networks of connected ``arms'', which are dynamically arrested in far-from-equilibrium states.  {Such} 
gels exhibit complex phenomena that continue to resist scientific understanding, including complex aging behaviour, and the possibility of self-induced catastrophic failure~\cite{zaccarelli2007colloidal,royall2021real}.  Perhaps surprisingly, this complexity can appear in systems with very simple ingredients, such as ``sticky spheres'' -- symmetric particles with short-ranged attractive forces.  While the equilibrium properties of such systems are well-understood~\cite{baxter1968,noro2000}, it is a challenging task to predict and control the properties of their non-equilibrium 
arrested states, including gels~\cite{segre2001glasslike,sedgwick2005non,rouwhorst2020nonequilibrium}.

{For example, it is understood in broad terms~\cite{lu2008gelation,dinsmore2002direct,shao2013role,tsurusawa2019direct,royall2021real, boromand2017structural,patrick2008direct,trappe2001jamming,bantawa2022hidden} that
the main control parameters for colloidal gels are the strength of attractive interactions between particles, and the particle volume fraction. 
These}  
parameters influence the mechanical properties of the gel, including its elastic response to small mechanical perturbations.  
When larger forces are applied, the arms of the gel network break, leading to macroscopic flow~\cite{poon2002physics,zaccarelli2007colloidal,trappe2004colloidal,cipelletti2005slow,masschaele2009direct,
gibaud2010heterogeneous,sprakel2011stress,grenard2014timescales,landrum2016delayed,gibaud2016multiple,johnson2018, koumakis2015tuning, nicolas2018deformation,cho2022yield}.
An important signature of the non-equilibrium gel state is that its
properties -- such as elastic moduli and yield stress -- depend on its history, including its age~\cite{fielding2000aging, zia2014micro,nabizadeh2021life,bartlett2012sudden, patinet2016connecting, parley2020aging, pollard2022yielding}.

The yielding of colloidal gels is important for practical applications~\cite{zaccarelli2007colloidal,royall2021real}, but it also raises fundamental questions for soft-matter modelling.  It
is sensitive to physical behaviour on several length scales \cite{richard2020}, from individual particle diameters \cite{barrat2018elasticity}, to the arm thickness, up to the size of a macroscopic gel \cite{lindstrom2012structures,colombo2014stress,verweij2019plasticity,bantawa2022hidden}.  There is a corresponding range of time scales \cite{song2022microscopic}: in large gel samples, there may be a long period of aging, after which the gel quickly collapses~\cite{bartlett2012sudden}.  This situation makes gels difficult to study.  Computer simulations can follow the motion of the colloidal particles, but the computational cost of simulating the long aging period is usually prohibitive, {especially given the large (macroscopic) system sizes that are required}. Experiments suffer from similar problems: techniques for monitoring a macroscopic sample do not allow individual particles' motion to be resolved, while particle-resolved techniques are limited to moderate time scales and cannot simultaneously resolve the behaviour throughout a macroscopic sample.  For theoretical analysis, continuum modelling approaches are available, but these do not resolve behaviour of individual particles, relying instead on constitutive models.  

This situation calls for an {approach on multiple scales}, in order to combine the useful aspects of different methods.  
In this work, we focus on the breakage of an individual arm (or strand), as a fundamental process that feeds into macroscopic yielding~\cite{sprakel2011stress}, and into aging (or coarsening) of gels~\cite{testard2011influence,Testard2014}.
{Like many processes in soft materials, this failure is challenging to predict and control, even with state-of-the-art theories: For example, the thickness of the arm ranges from $1$ to $10$ colloidal diameters, so single-particle fluctuations can have significant impact on the whole arm: this limits the predictive power of continuum models.  Also, the strength of attractive interactions is only a few times the thermal energy, so there are frequent fluctuations in which interparticle bonds are broken and re-formed, necessitating a statistical mechanical approach.  In addition, the arm is itself an amorphous material, so subtle features of the particle-level structure can have significant effects on its large-scale behavior, as is familiar in glassy materials.}

\begin{figure*}[t] 
    \centering
    \includegraphics[width=0.95\textwidth]{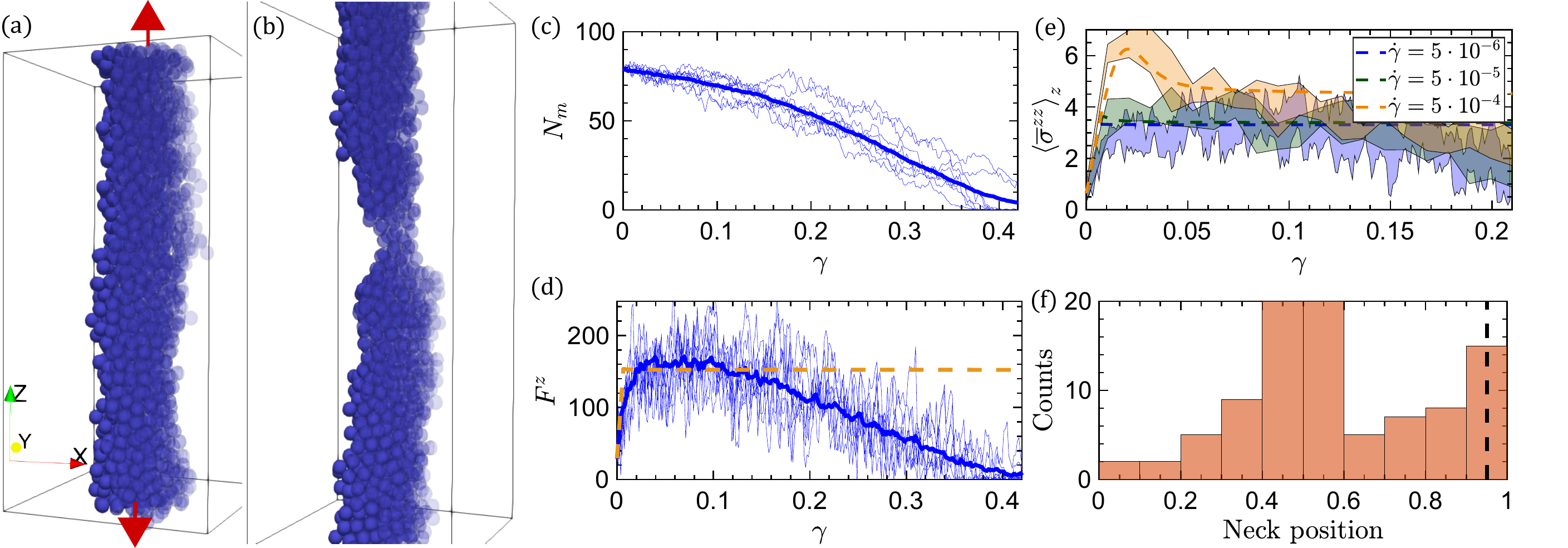}
    \caption{
(a) A gel strand after initialization $(\gamma=0)$ with uniform thickness, red arrows show the direction of elongation.  Parameters: $\epsilon=4.5$, $r_0=4 l$ and $\dot{\gamma}=5\times10^{-6}$.
(b) The same strand at strain $\gamma=0.4$, showing formation of a neck.
(c,d) The number of particles in the thinnest region of the strand $N_m$ and the tensile force in the strand $F^{z}$, as a function of strain $\gamma$. 
Thin lines show 8 runs from the same starting conditions (thin blue lines). Thick lines are an average over 100 runs.
(e) The cross-sectional stress $\overline{\sigma}^{zz}$, averaged over $z$, for different varying strain rates $\dot{\gamma}$. (Shaded regions indicate the standard deviation over 6 separate runs).  The dashed lines in (d,e) are homogeneous solutions of the continuum theory, which are valid for times before significant necking has occurred (see main text).
(f)  Distribution of the position of the neck, for 100 runs with identical initial conditions. Dashed line indicates the thinnest segment at $\gamma=0$. 
}
\label{fig:multiple_runs}
\end{figure*} 

{In response to these challenges, we performed computer simulations of the failure of a single arm, under elongation. We analyse the results
using three complementary methods: a simple continuum modelling approach~\cite{hoyle2015age,hoyle2016criteria,moriel2018necking}; an analysis at single-particle level; and measurements of the local stress~\cite{irving1950statistical}.
Failure occurs by a necking mechanism, which
proceeds via a feedback mechanism, leading to a linear instability.  Such instabilities may generally occur in several different ways~\cite{hoyle2015age,hoyle2016criteria,moriel2018necking}: Our results show that the tensile force in the strand generates an increased  stress in the neck, resulting in increased plastic flow there, which causes further thinning. This increases the local stress even more, and so on, until the strand breaks.  In simple terms, one may imagine that the stress $\sigma$ exceeds the yield stress $\sigY$ in the neck} \cite{barrat2018elasticity}, while remaining below $\sigY$ elsewhere.  Plastic flow in the neck is revealed by local measurements of increased particle motion, and this is coupled with a reduction in the number of low-energy (stable) structures.  We refine this picture in two ways.  First, a top-down continuum modelling approach~\cite{moriel2018necking} indicates the existence of an internal {time-dependent} plasticity field that determines the response to local stress.  Second, particle-level measurements of the local stress~\cite{irving1950statistical} reveal a complex pattern of residual stresses \cite{tsamados2008study,tsamados2009local, vinutha2022stress} that come from the initialisation of the arm \cite{zhang2022prestressed}; this pattern changes significantly in the neck, as it develops.  {This is reminiscent of other amorphous systems where mechanical heterogeneous properties can determine plasticity \cite{yoshimoto2004mechanical, mizuno2013measuring, tsai2017hierarchical,bian2019controlling}.}
  These results bridge the scales between continuum modelling (rheology) and the particle level (local motion and local structure), allowing new relationships to be revealed.  The connections are mediated by our direct measurements of local stress.

By combining these analyses on different scales, our results greatly extend previous work on single gel strands~\cite{van2018strand,verweij2019plasticity} and on breakage of glassy (amorphous) samples~\cite{moriel2018necking,richard2022arxiv}.
Some other approaches to gel modelling~\cite{brambilla2011,secchi2014} consider the gel as a continuum, without resolving individual arms -- our approach can connect such models with the microscopic gel structure.
{Similarly, our {detailed analysis of} 
individual arms complements alternative simulation approaches where
 particle interactions are justified by a top-down approach~\cite{bouzid2017elastically,bantawa2022hidden}: these models do not resolve the microscopic structure of the arms, but they do enable simulations of an entire gel sample.}

{Overall, our results elucidate the microscopic mechanism for  strand failure. They include the discovery of new relationships between emergent properties (like yielding) and local (microscopic) structure.  An understanding of these relationships, which can serve as a foundation for the coarse-grained modelling of macroscopic gels,  is vital for the design and control of gels, as an important class of soft materials.}

\section*{Results}

\subsection{Overview}
\label{sim}

We briefly describe our model system, with full details in Methods.  We simulate a gel-forming system~\cite{taffs2010,razali2017effects}  of particles with short-ranged attractive interactions, and two different particle sizes, to suppress crystallisation
\cite{razali2017effects}.
Particle $i$ has position $\vec{r}_i$ and velocity $\vec{v}_i=\dot{\vec{r}}_i$, it evolves by Langevin dynamics in a simulation box with periodic boundaries: 
 \begin{equation}
m\dot{\vec{v}}_i=-\nabla_i \mathcal{V}-\dampbare \vec{v}_i+ \vec{F}_{\rm solv} \; ,
\label{equ:eom}
\end{equation}
where $m$ is the particle mass, $\mathcal{V}$ is the interaction energy, $\dampbare$ is a friction constant, and $\vec{F}_{\rm solv} $ the random solvent force.
The mean particle diameter is denoted by $l$. 
We initialise the system to mimic a single arm of a colloidal gel, by quenching a bulk colloidal liquid to form an amorphous (glassy) solid, and excising a cylindrical sample,  see Fig.~\ref{fig:multiple_runs}(a).  The arm has initial length $L_{\parallel}=30l$ and we vary its initial radius $r_0$ to mimic arms of different thicknesses.  {An important time scale is the Brownian time $\taub$ (see Methods), which is the typical time required for an individual particle to diffuse its own radius.}

We deform the arm by affine elongation of the simulation box, so the length $L_\parallel(t)$ increases with time $t$.  This stretches the arm,  which eventually breaks.
Fig.~\ref{fig:multiple_runs}(a,b) illustrates the resulting behavior, which is governed by 4 dimensionless parameters (see Methods): 
the attraction strength $\epsilon$ between the colloids, a rescaled elongation rate $\dot\gamma$, the arm thickness $r_0/l$, and a solvent damping parameter $\damp$.
{Gelation is associated with metastable colloidal gas-liquid phase separation \cite{royall2021real}, whose critical point is at $\epsilon^*\approx 2.8$ \cite{noro2000}.  The data of Fig.~\ref{fig:multiple_runs} have $\epsilon=4.5$, corresponding to $\epsilon/\epsilon^*\approx 1.6$, well inside the spinodal.} We fix $\damp=10$ throughout, consistent with a colloidal gel in a high-friction solvent environment.
For the other parameters, we mostly focus on the representative values used in Fig.~\ref{fig:multiple_runs}(a-d).
The qualitative picture is robust to changing these parameters, this will be discussed below.

Figure~\ref{fig:multiple_runs}(b) illustrates the neck that forms as the arm is stretched.  It becomes increasingly thin and eventually pinches off, so the arm breaks.
To focus on this effect, we divide the simulation box into 20 segments along the $z$-direction, indexed by their rescaled positions $Z=z/L_\parallel(t)$.  
We count the number of particles in each segment, and write $N_m$ for the smallest such number, which serves as a proxy for the thickness of the neck.
Fig.~\ref{fig:multiple_runs}(c) shows $N_m$, as the arm is elongated.  To illustrate the variability in the failure mechanism, we show results for 8 representative simulation runs, all from the same initial structure, with different random forces, (such isoconfigurational ensembles have been used before in studies of glassy materials~\cite{widmer2007study}).  We also show the average of $N_m$, obtained over 100 such runs.  The formation of the neck is clear, as is the gradual reduction in its thickness, leading to failure.

We measure the stress in the arm by the method of Irving and Kirkwood (IK)~\cite{irving1950statistical,yang2012generalized,smith2017towards}.   
The resulting stress estimates are noisy, due to rapid thermal fluctuations, so all such measurements are averaged over a time period of $250\taub$, to reduce the statistical uncertainty.
The effects of this averaging are discussed in Appendix~A.
Throughout this work, all stresses (and elastic moduli) are quoted in units of $k_{\rm B}T/l^3$.

The strand is under tension and the corresponding tensile force is obtained from the stress: its average is shown in Fig.~\ref{fig:multiple_runs}(d), together with several representative trajectories.  There are three clear regimes, as the system elongates.  Initially, the response is elastic and the stress increases.  This is followed by a short plateau -- a signature of plastic flow -- before the stress smoothly decreases, until breakage occurs.  This decrease mirrors the reduction in the neck thickness [Fig.~\ref{fig:multiple_runs}(c)].   
Physically, the force required to maintain a constant strain rate is decreasing, as the neck gets thinner.  However, there is no sudden drop in stress, as would be expected for brittle failure.  In all measured cases, the neck gets gradually thinner, until the entire tensile force has to be sustained by a cross-section containing only 1-2 particles.  At this point the tension is very small, and the strand ruptures.

In addition to the tensile force, the IK method also provides a local measurement of the stress, this is a $3\times3$ matrix which we denote by $\sigma(\vec{r})$.  
To make contact with continuum models of rheology~\cite{hoyle2015age,hoyle2016criteria,moriel2018necking}, we consider the $zz$ component of the stress, averaged over the cross-section of the arm (see Methods). The resulting quantity is denoted by $\sigbarz(Z)$.   After averaging this quantity over $Z$, we plot the result in Fig.~\ref{fig:multiple_runs}(e), for several different values of the elongation rate $\dot\gamma$.  These results are compared with a simple rheological model (dashed lines), see below for details.

Recall that we performed multiple simulation runs from the same initial configuration.  For this initial condition,
Fig.~\ref{fig:multiple_runs}(f) shows that the neck is more likely to occur in particular locations.  Simple theories of the necking instability~\cite{hoyle2016criteria,moriel2018necking} indicate that this location should be the thinnest part of the initial state -- this is not consistent with the data, which is a first indication that the internal structure of the arm is playing a role in the rheology.

\begin{figure} 
    \centering
    \includegraphics[width=0.45\textwidth]{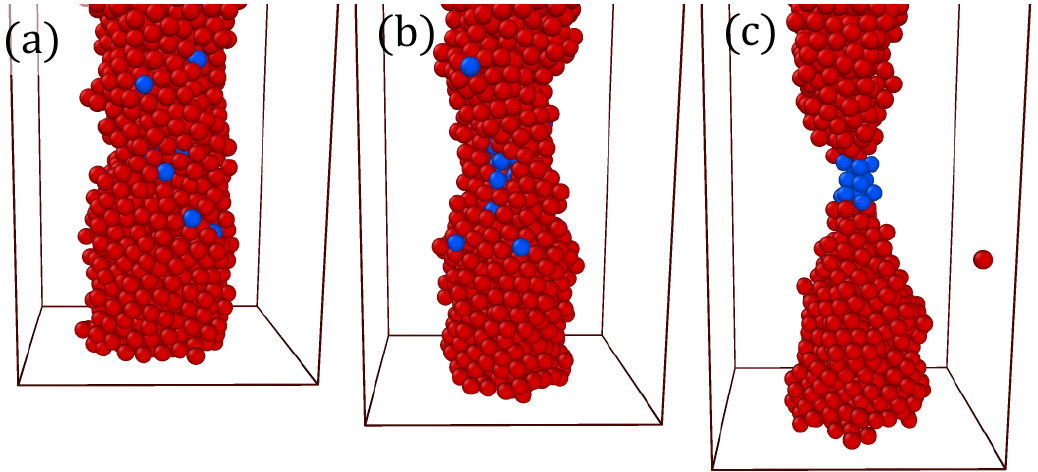}
    \caption{
    Snapshots of the the strand from a single trajectory, as the neck forms.  (a) Strain $\gamma=0.06$, (b) $\gamma=0.20$, (c) $\gamma=0.36$. The 16 particles that form the neck in (c) are coloured in blue, to show their movement.}
\label{fig:Details_failure}
\end{figure}

Before analysing these effects in more detail, we identify a surprising aspect of neck formation: Fig.~\ref{fig:Details_failure} shows that the particles in the neck just before breakage (coloured blue) have arrived in the neck region from a range of other locations in the arm.  
This demonstrates significant mobility of particles, especially for those near the surface of the strand, and for those in the neck.

\subsection{Continuum-scale description}

{We develop a simple description of the necking process in the continuum. On this scale, two distinct types of necking instability in amorphous cylindrical arms have been identified ~\cite{moriel2018necking, hoyle2015age,hoyle2016criteria}: gradual plastic deformation or sudden failure.
The behaviour found in our system corresponds to the plastic (gradual) case.}  
(The sudden mechanism -- not seen here -- is driven by the build up of elastic stress.)  

{Following~\cite{moriel2018necking, hoyle2015age,hoyle2016criteria}, we define}  $Z=z/L_\parallel(t)$, a rescaled co-ordinate along the arm. Modelling this arm as a thin filament, the theory is based on four $Z$-dependent quantities: the filament's cross-sectional area $a(Z)$; the local strain rate $\dot\gamma_{\rm L}(Z)$; a scalar field $W(Z)$ that coincides (in this case) with the stress $\overline{\sigma}^{zz}(Z)$; and a plasticity field $\chi(Z)$ that controls the plastic flow rate {(see below)}. 

The elongation rate is slow enough that advective effects are negligible during elongation, and local force balance holds at all times. Then
  conservation of mass requires $\partial_t a=(\dot{\gamma}-\dot{\gamma}_L)a$, and the force balance condition is is $\operatorname{div}\sigma=0$ (see Methods). For the filament, this means that 
\begin{equation}
\frac{\partial}{\partial Z}(a W)=0 \; .
\label{equ:force-bal}
\end{equation}
We identify $aW$ as the tensile force in the arm, which is constant along its length.

\begin{figure*} 
    \centering
    \includegraphics[width=0.94\textwidth]{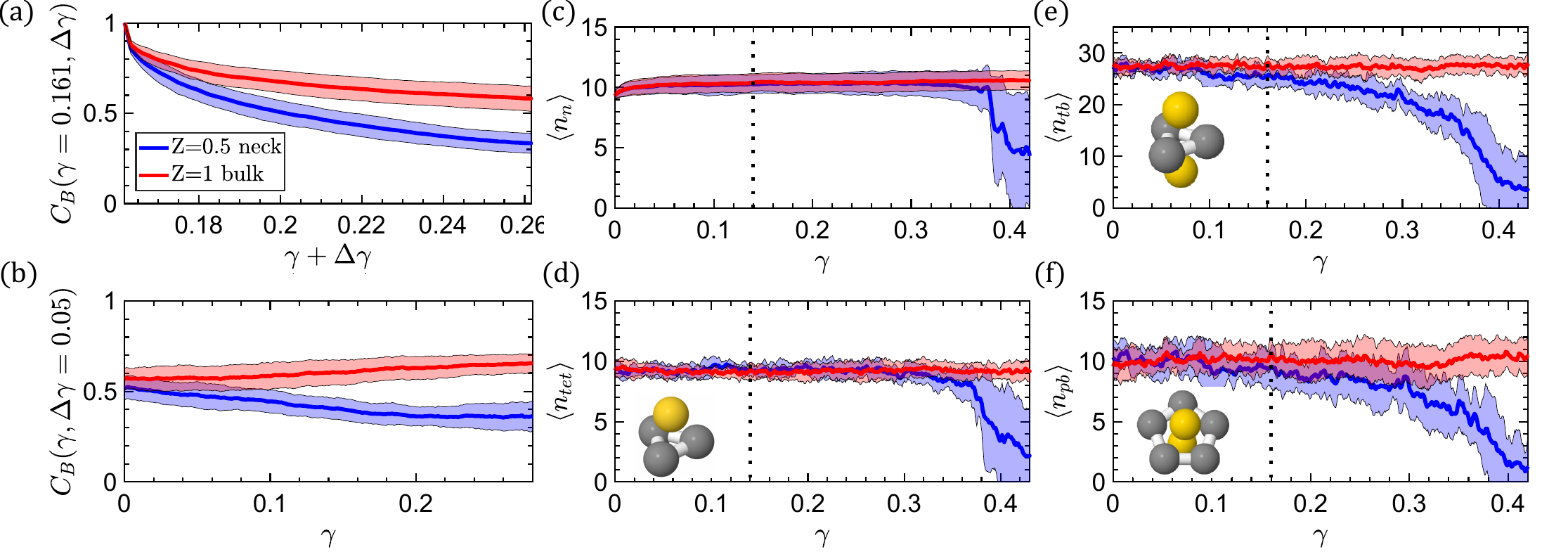}
    \caption{
    (a) Dynamical correlation function $C_B(\gamma,\Delta\gamma)$ on varying $\Delta\gamma$, comparing the neck ($Z=0.5$) and the bulk ($Z=1$).  The correlation function decays as the particles move away from their neighbours.
    (b) Dynamical correlation on varying $\gamma$ at fixed lag $\Delta\gamma$.  Increasing separation of the curves indicates dynamical contrast between the neck and the bulk.
    (c) The ensemble-averaged number of neighbours $n_n$.
    (d-f) The ensemble-averaged number of tetrahedra $n_{\rm tet}$, triangular bipyramids $n_{\rm tb}$ and pentagonal pipyramids $n_{\rm pb}$ in which a particle participates. 
    The legend in (a) is common to all panels.
     Data was averaged over 10 trajectories, parameters are the same as shown in fig \ref{fig:multiple_runs}. Shaded regions indicate the standard deviation.
    The dotted line indicates the onset of necking (the strain beyond which $N_m$ has decreased significantly).
   }
\label{fig:CN}
\end{figure*} 

{Our equation of motion for $W$ assumes that stress increases due to elastic loading and relaxes by plastic flow:
\begin{align}
\frac{\partial W}{\partial t}= G\left[\dot{\gamma}-p(W,\chi)\right],   \label{eq:W_stress}
\end{align}
where $G$ is the elastic modulus and $p$ is the rate of plastic relaxation~\cite{moriel2018necking}, which depends on the stress and the plasticity $\chi$.  We assume that plastic flow only takes place above the yield stress $\sigY$, taking
$p(W,\chi)=\tau_p(\chi)^{-1}
\left[(W/\sigY)-1\right] \theta \left(W-\sigY\right)$ where $\tau_p$ is a plastic time scale and $\theta$ is the Heaviside (step) function.
Following~\cite{moriel2018necking}, the plasticity $\chi$ is analogous to a local temperature that determines the probability of an an activated rearrangement, inspired by the theory of shear transformation zones~\cite{falk1998}.  Hence, $\tau_p(\chi) = \tau_{\rm ref}\exp(1/\chi)$ where $\tau_{\rm ref}$ is a reference time scale.}

{The general approach of~\cite{hoyle2015age,hoyle2016criteria}
starts with homogeneous solutions to the equations of motion (that is, $a,\dot\gamma_L,W,\chi$ independent of $Z$).}  Necking is a linear instability of this solution.  For the homogeneous solution, we first assume that the field $\chi$ is a simple constant, independent of both $t$ and $Z$.  One finds $W(t)=G\dot\gamma t+ W(0)$ for short times (such that $W<\sigY)$ while for long times $W(t)=\sigY/(1-\dot{\gamma}\tau_p)$.  This theory was used to fit the data of Fig.~\ref{fig:multiple_runs}(d), in the early-time regime before the neck forms.  However, the data for faster elongation rates in Fig.~\ref{fig:multiple_runs}(e) shows a stress overshoot, which cannot be described at this level of theory.  

To account for this effect, we allow the plasticity field $\chi$ to depend on time with a relaxational dynamics that is controlled by the plastic relaxation itself~\cite{moriel2018necking}:
\begin{align}
\frac{\partial \chi}{\partial t}=\frac{W(t)}{\sigY}p(W,\chi) \left[\chi_\infty-\chi(t)\right].
\label{eq:chi}
\end{align}
The additional parameter $\chi_\infty$ fixes the steady-state value of $\chi$.  

This model supports two main regimes.  For very slow elongation with $\dot\gamma  \tau_p \ll 1$, the field $\chi$ is constant until $W$ reaches the yield stress, at which point $\chi$ relaxes quickly to $\chi_\infty$.  The slowest elongation rate in Fig.~\ref{fig:multiple_runs}(e) is consistent with this regime, fitting gives $\sigY \approx 3.3$ (recall the units are $k_{\rm B}T/l^3$).  For faster elongation rates, the relaxation of $\chi$ competes with the elongation rate, this is responsible for the stress overshoot in Fig.~\ref{fig:multiple_runs}(e).  Fitting simultaneously to all the curves in \ref{fig:multiple_runs}(e), we estimate $G=420$, $\chi_\infty=80$ and $\tau_{\rm ref}=180\taub$, as well as the initial condition $\chi(t=0)=0.65$, see Appendix for further details.  We also note that while the model proposed here fits the data satisfactorily, fits of similar quality can also be achieved with different choices for $p(W,\chi)$, and different dynamics in Eq.~\ref{eq:chi}.  The essential physical features that are required to fit the data are the existence of a yield stress (which ensures that the stress plateau in Fig.~\ref{fig:multiple_runs}(e) depends weakly on $\dot\gamma$), and the non-trivial time dependence of $\chi$ (which accounts for changes in structure on elongation, and allows the model to capture the stress overshoot).

Having characterised the homogeneous solutions for $W$, we can now perform a linear stability analysis by allowing Z-dependent perturbations, see Appendix.
The results indicate that the system is always unstable once $W$ rises above $\sigY$, so necking should occur in all cases, as observed.   

A simple physical picture of this instability is that since the tensile force is constant along the strand [Eq.~\eqref{equ:force-bal}], the stress is largest at its narrowest point: this will be the location where $W$ first exceeds $\sigY$, leading to plastic flow near this point, and hence to further thinning.  However the observation  of Fig.~\ref{fig:multiple_runs}(f) -- that the neck does not always form at the thinnest point -- indicates that heterogeneities in the internal structure of the arm also play a role in the instability.  These features might enter the model as inhomogeneities in $\chi$, {whose analysis requires that we go beyond the thin-filament description of the arm, and consider its internal (microscopic) structure.} 

\subsection{Particle-level description}

\begin{figure*}
    \centering
    \includegraphics[width=0.98\textwidth]{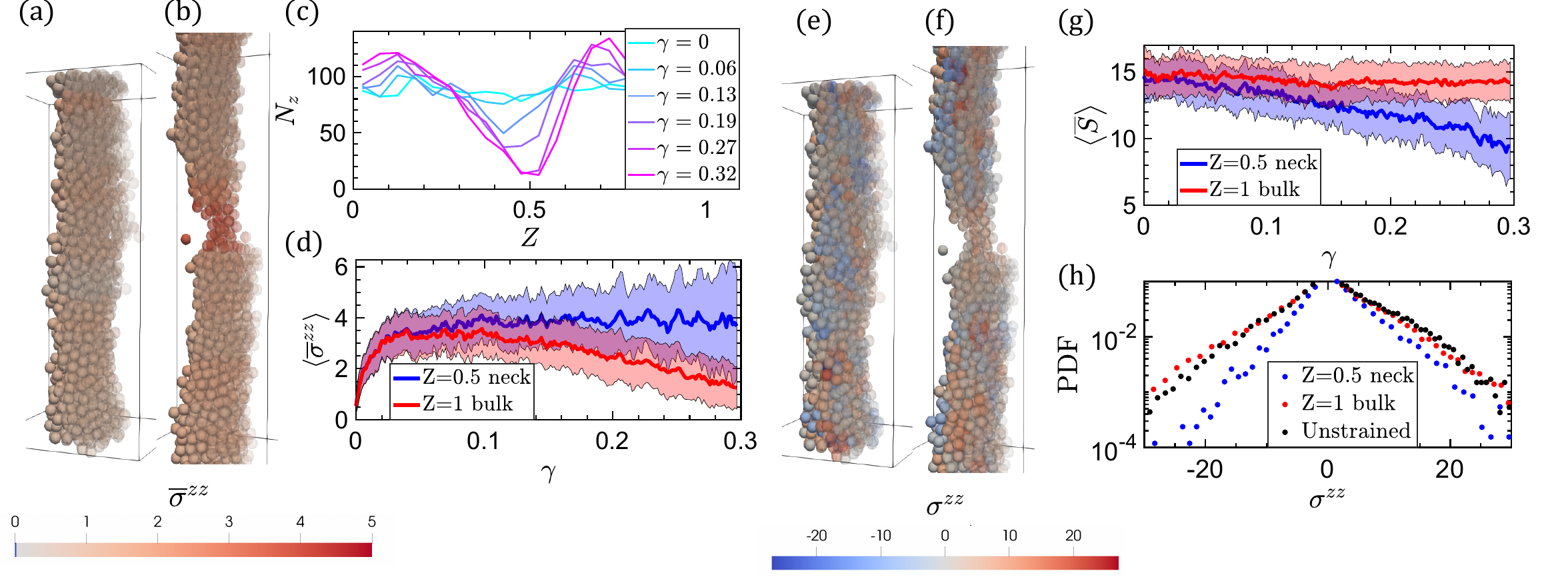}
    \caption{
   (a) The $Z$-dependent stress $\overline{\sigma}^{zz}$ at initialization. (b) The stress at $\gamma=0.4$.
    (c) The number of particles in different segments of the arm, as the strain increases, for the single trajectory shown in (a,b).
    (d) The average  stress profile $\overline{\sigma}^{zz}(Z)$ at the neck $(Z=0.5)$ and in bulk $(Z=1)$ as a function of strain $\gamma$.  (Data averaged over 6 runs.)
    (e,f) The local stress $\sigma^{zz}$, for the same configurations shown in (a,b).
(g)  The anisotropic stress measurement $\overline{S}(Z)$ as a function of strain $\gamma$.
(h) The stress distribution curves just before failure ($\gamma = {0.35-0.4}$) for the neck and the bulk, compared with the unstrained distribution. 
    }
\label{fig:Macro}
\end{figure*} 

Having analysed the behaviour at the level of the entire arm, we now turn to the microscopic (particle-level) structure.
Recalling that the simulation box is divided into 20 segments, we identify for each run the segment in which the arm eventually breaks.  We then make a transformation of spatial coordinate so that this segment corresponds to $Z=0.5$ (the centre of the simulation box).  We compare the physical properties of this central cell (``neck region'') with those at $Z=1$ (``bulk region'', far from the neck).  The results are obtained by averaging over 100 trajectories, all for the same initial condition.  (We checked that the same behaviour occurs also for other initial conditions.)

We begin with a comparison of dynamical quantities.  We define a correlation function $C_{\rm B}(Z;\gamma,\Delta\gamma)$ that measures how much particles' local environments have relaxed in segment $Z$, for the period between strain $\gamma$ and strain $\gamma+\Delta\gamma$ (see Methods).

Figure~\ref{fig:CN}(a,b) shows that particles in the neck region have significantly faster relaxation than those in the bulk.  
This is characterised in two ways: Fig.~\ref{fig:CN}(a) fixes $\gamma$ at a value where the neck has already formed, showing that $C_{\rm B}$ decays faster for the neck and slower for the bulk, as a function of $\Delta\gamma$.
On the other hand, Fig.~\ref{fig:CN}(b) fixes the time lag $\Delta\gamma=0.05$ and varies $\gamma$, showing how the difference in relaxation rate grows, as $\gamma$ increases and the neck develops.  This is a microscopic signature of the prediction of the continuum theory, that plastic flow events occur preferentially in the neck.

We also analyse the local structure of the neck,
using the topological cluster classification (TCC) \cite{malins2013tcc}.  This provides a detailed characterisation of local packing, by identifying specific geometrical structures within the system.  We write $n_{\rm n}(Z;\gamma)$ for the average number of neighbouring particles around a particle, at position $Z$ and strain $\gamma$.  Similarly we write $n_{\rm tet}$ for the average number of fully-bonded tetrahedra in which a particle participates, and  $n_{\rm tb}$ and $n_{\rm pb}$ for numbers of triangular and pentagonal bipyramids \cite{jack2014information}.  (These local packing motifs are illustrated in Fig.~\ref{fig:CN}.)

Figure~\ref{fig:CN}(c-f) shows these {quantities} 
for both the neck and the bulk, as $\gamma$ increases.   The average {number of neighbours} 
$n_{\rm n}$ and {the average number of tetrahedra $n_{\rm tet}$ in which a particle participates} remain almost constant with $\gamma$, until the system starts to fail around $\gamma=0.36$.  (The large error estimates in this region arise because failure occurs at different values of $\gamma$ for different runs.)  Both these quantities follow the number of neighbours, which means that the neck differs significantly from the bulk only when the system is close to failure.   

{However, we find different behaviour for
the average numbers of triangular bipyramids $n_{\rm tb}$ and pentagonal bipyramids $n_{\rm pb}$ in which a particle participates.
These measurements correspond to larger structures of 5 and 7 
particles respectively, which are
sensitive to details of the packing, 
and tend to be more common in amorphous materials that are deep in the energy landscape.  Hence $n_{\rm tb}$ and $n_{\rm pb}$ are larger in materials that are well-annealed and stable.} 
The neck shows a deficit in these low-energy structures, compared with the bulk (where the number is almost constant): these differences appear as the neck begins to form 
($\gamma\approx0.1$).  
This effect can be understood in terms of the faster dynamics in the neck, which tends to break up low-energy local structures, as happens in sheared bulk samples~\cite{ding2014,pinney2016,pinney2018,richard2020}.

\subsection{Local stress}

In order to bridge scales between the particle-level and  the whole arm, we measure the local 
stress, which is the fundamental object for controlling rheology.  

The stress $\overline{\sigma}^{zz}(Z)$ corresponds to the tensile force in the arm divided by its cross-sectional area, at position $Z$.  This quantity was already discussed in Fig.~\ref{fig:multiple_runs}(e), which shows its behaviour after averaging over $Z$.  The $Z$-dependence of this quantity is shown in Fig.~\ref{fig:Macro}(a,b): one sees that the formation of the neck is accompanied by a local increase in stress.  This result was already anticipated in Eq.~(\ref{equ:force-bal}) and the discussion at continuum level. (Physically, the tension is constant along the arm, so the stress must be larger in segments where the cross-sectional area is smaller.)  Fig.~\ref{fig:Macro}(c) shows the development of the neck, and Fig.~\ref{fig:Macro}(d) compares the stress in the neck with the stress in ``the bulk'' (far from the neck).  The central observation is that the stress in the neck reaches a plateau at the yield stress of the arm, so plastic events continue to occur there, allowing the arm to elongate.  As the neck develops, the reduction in area at (almost) constant stress explains the reduction in the tensile force [Fig.~\ref{fig:multiple_runs}(d)].  This decreasing force leads to a reduction in the bulk stress, which falls below $\sigY$.  Hence these measurements show directly the mechanism by which plastic events concentrate in the neck.

\begin{figure*} 
    \centering
    \includegraphics[width=0.98\textwidth]{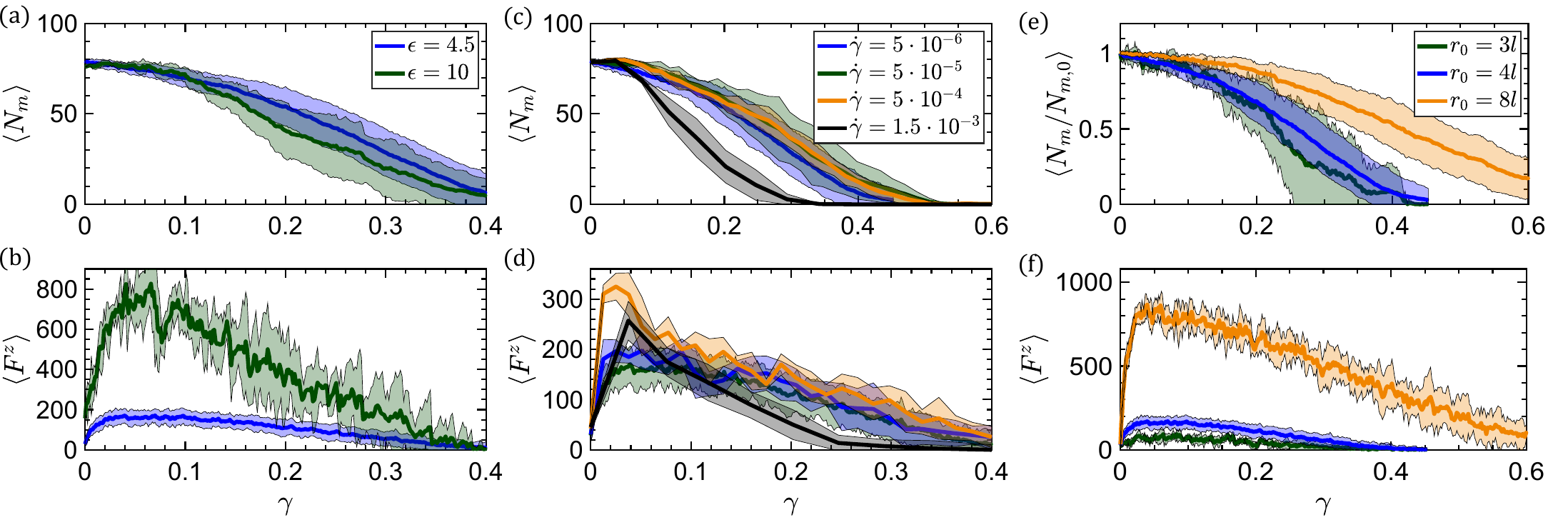}
    \caption{
    Dependence on model parameters.  (a,c,e) show the average neck thickness and (b,d,f) show the tensile force in the arm.  In (a,b)  the attraction strength $\epsilon$ is varied; in (c,d) it is the elongation rate $\dot\gamma$; and in (e,f) it is the arm radiue $r_0$.
All lines are averaged over 6 runs with identical starting conditions for a given set of parameters. 
The blue lines correspond to baseline parameters: $\epsilon=4.5$, $r_0=4 l$ and $\dot{\gamma}=5\times10^{-6}$.
}
\label{fig:strain_stress}
\end{figure*}

A more detailed analysis of the stress reveals additional complexity.  Fig.~\ref{fig:Macro}(e,f) shows the $zz$ component of the local stress $\sigma^{zz}$, averaged over small regions of size $(1.5l)^3$.  Before elongation, one might have imagined that the local stress in the arm [Fig.~\ref{fig:Macro}(e)] would be close to the average behaviour [Fig.~\ref{fig:Macro}(a)].  Instead, $\sigma^{zz}$ has strong inhomogeneities: its average is positive (corresponding to tension), but there are significant regions where $\sigma^{zz}<0$.
 These inhomogeneities are residual stresses, arising from the (non-equilibrium) process of initialising the arm.
We verified that the resulting spatial structures are long-lived and tied to the amorphous structure of the material (see Appendix), they are not the result of fast thermal fluctuations.  

We also checked that the arms satisfy local force-balance ($\operatorname{div}{\sigma}=0$), the residual stresses are divergence-free and arise from the inhomogeneous (amorphous) microstructure of the arms.  Since $\sigma^{zz}$ depends quite strongly on $z$, this requires that the stress also has significant off-diagonal components.
Such residual stresses were not included in the simple rheological models considered above, where the stress $W$ depended only on $Z$.  
However, bridging between rheological models and the local motion of individual particles requires some analysis of these stresses.
For example, the {plasticity} $\chi$ presumably depends on the local structure of the arm and the residual stresses there -- this offers the opportunity for a deeper understanding of the physical meaning of this field, if it could be connected to particle-level observables.

To explore this, Fig.~\ref{fig:Macro}(f) shows the behaviour as the arm is elongated and the neck forms.  An interesting feature is that the local stress in the neck region appears more homogeneous, compared to the bulk.
To quantify this observation, we measure the anisotropy of the stress tensor $\sigma$:  define 
$\bar S(Z)$ by averaging the (square root of the) second stress invariant over a given segment of the arm (see Methods).  
This quantity tends to be large when the stress has large fluctuations away from its average, within a segment; it is smaller when the stress is homogeneous.  
After averaging over many trajectories, Fig.~\ref{fig:Macro}(g) compares $\bar S(Z)$ in the neck with the bulk of the strand.  It shows that the neck does indeed have a more homogeneous stress field.  

Also, Fig.~\ref{fig:Macro}(h) shows the distribution of $\sigma^{zz}$, within the neck and the bulk.  One sees that the probability to see $\sigma^{zz}<0$ is suppressed in the neck.  In fact, it seems that the reduction in this probability accounts for most of the increase in the average $\sigbarz$ in the neck (the typical value of $\sigma^{zz}$ barely increases there).  
In comparison with these results, we also simulated the elongation and rupture of crystalline arms (details are given in the Appendix).  In this case,
the residual stresses are (mostly) absent and there is no homogenisation of the stress within the neck [for example, $\bar S(Z)$ does not decrease there.]  

Coupled with Fig.~\ref{fig:CN}, these results show that plastic flow in the neck is coupled with a two distinct local changes: reduction in low-energy (stable) structures, but also a homogenisation of the residual stresses [smaller $S(Z)$].
Physically, we can think of this process as a partial fluidisation of the neck, where it loses its amorphous-solid character and starts to move as a highly-viscous fluid, {under the applied stress}.  The solid is characterised by stable local structure and large residual stresses; the fluid is more disordered in its structure but more homogeneous in stress.  {Similar behavior has been observed in metallic glasses in the spreading of shear transformation zones \cite{csopu2017atomic, bian2020signature}}. 
Overall, these results illustrate the subtle interplay of structure, forces, and dynamics, which is required to characterise the multi-scale phenomenon of yielding and fracture in these amorphous systems.

\subsection{Robustness of results for different model parameters}

The results presented so far focussed on representative parameter values: attraction strength $\epsilon=4.5$, strain rate $\dot{\gamma}=10^{-6}$ and strand thickness $r_0=4l$. However, the general picture that we present is robust.  This is shown in Fig.~\ref{fig:strain_stress} where we vary the strain rate $\dot\gamma$, the arm thickness $r_0$, and the cohesive energy $\epsilon$.  

Varying the strength of attractive forces  [Figs.~\ref{fig:strain_stress}(a,b)], there is weak dependence of the neck thickness $N_m$ on $\epsilon$.  The tensile force in the arm is larger when the attractive forces are stronger (as expected), but the qualitative behaviour of the stress remains the same.  (This is also true for the local stress.)  We did find that for larger $\epsilon$, the position of the neck formation was more predictable [the distribution analogous to Fig \ref{fig:multiple_runs}(f) was more sharply-peaked, data not shown]. 

The effect of varying elongation rate is shown in Figs.~\ref{fig:strain_stress}(c,d).  It is notable that the neck thickness $N_m$ is almost unchanged as $\dot\gamma$ varies over two orders of magnitude.  For yet larger strain rates, the thinning process of the neck is accelerated but the qualitative picture remains the same.  
Fig.~\ref{fig:multiple_runs}(e) already showed that the stress during neck formation depends weakly on the strain rate, although faster elongation does create a stress overshoot.  This picture is confirmed by Fig.~\ref{fig:strain_stress}(d) which shows the corresponding tensile force.  We also performed simulations for even faster elongation: the picture of  a gradual (viscosity-driven) necking instability remains robust and we did not observe any sudden (elasticity-driven) failure mode~\cite{eastgate2003dynamics,hoyle2015age,hoyle2016criteria}, {nor solid--like brittle failure.}

Finally, varying the arm thickness $r_0$  [Fig.~\ref{fig:strain_stress}(e,f)] one sees a general trend that thicker arms fail at larger strains (and support somewhat larger stresses), but the qualitative behaviour is again the same.

\section*{Discussion}

{We have addressed some of the challenges posed by failure of colloidal gels, by operating at different scales. By analysing a single arm, we are able to implement a continuum description, while we also analyse our simulations at a single--particle level using higher--order structure, and also at a more coarse--grained level with the stress tensor.}

At the level of the whole strand, we find that failure occurs by a necking instability, driven by plastic flow.  This can be captured by a simple continuum model, following~\cite{moriel2018necking,hoyle2016criteria}. 
{This description of the necking will be a useful basis for future models of failure in these challenging heterogeneous materials.}

At the microscopic level, we showed that particle dynamics are faster in the neck, and that {complex higher--order structures associated with rigidity are signficantly suppressed prior to failure. Remarkably, the structural signatures prior to failure appear to be rather stronger than those in glasses undergoing shear--induced failure \cite{ding2014,pinney2018,turci2018}.}

To bridge between these levels, we analysed the local stress: this is a fundamental quantity for the rheology, which can also be analysed at the microscopic level.
Our results demonstrate links between the local structure and the stress: the structural and dynamical changes in the neck are accompanied by a reduction in the residual stresses, which makes the stress field more homogeneous.

Our microscopic resolution of the stress provides mechanistic insight, in that the stress $\sigbarz(Z)$ has a value that remains close to the yield stress $\sigY$, as the neck develops [Fig.~\ref{fig:Macro}(d)].  The constant tensile force in the arm means that other (thicker) parts of the arm have $\sigbarz(Z)<\sigY$, so that plastic events are increasingly concentrated in the neck.  This positive feedback drives the (plastic) necking instability, and the mechanism is consistent with the simple continuum model.

On the other hand, the simulations reveal large  residual stresses in the arm, which are not accounted for directly in the continuum approach.  These results reinforce the observation that the amorphous structure of the arm is heterogeneous, both in its local structure and in the stress field.  Consequences of this heterogeneity include Fig.~\ref{fig:multiple_runs}(f), which shows the place where the arm is most likely to break.  This issue -- of predicting rupture -- would seem to be vital for predicting and controlling properties of gels.  Such issues are not easily addressed by continuum modelling,  their resolution will require input from particle-level data.  For example, a simple von Mises criterion for yielding in soft solids~\cite{timoshenko1956strength,sica2020mises} predicts that a sample will fail in places where the local stress is large and anisotropic: the residual stresses mean that such locations do exist in our samples, but we do not find them to be correlated with the failure.  

An alternative approach -- building on the continuum theories -- would be to search for links between the continuum-level plasticity field $\chi$ and the stress anisotropy (or local structure) at the microscopic level.  It is certainly true that some aspects of the microscopic structure should influence the continuum models through such a field, as well as through sample-to-sample variations in constitutive parameters such as the yield stress and elastic modulus.  These issues would be usefully investigated in future work.

In addition to these questions about the failure of a single arm, our results can also inform future studies of gel behaviour.  In particular, we imagine extending these measurements to gel samples under shear~\cite{johnson2018}, or to analyse the breakage of arms during coarsening~\cite{Testard2014}.  In these cases, the gel will have many arms, and its overall failure will depend on which ones break first, and on how aging affects the tendency to failure.  For example, the results here indicate that the more fluid behaviour of the strand in the neck region might be a useful early-warning signal of strand breakage.  Such signals would be helpful for characterising and predicting the behaviour of gels as they undergo ageing and/or collapse.
{An important extension of this work would be  {a detailed comparison of} our results with experiments, for example {rheological studies}~\cite{laurati2011} 
and {confocal microscopy observations}~\cite{smith2017towards}.
Finally, we note recent work where imaging has enabled forces between colloidal particles to be inferred, which could provide a direct comparison of the stress field \cite{dong2022direct}.}

\section*{Methods}

\subsection*{Simulation model and dimensionless parameters}

We consider a bidisperse system of $N$ colloidal particles, the two species have diameters $1.04l$ and $0.96l$, where $l$ is the average size.  They interact by a Morse potential, which mimics the short-ranged depletion interaction \cite{taffs2010,razali2017effects}
\begin{equation}
    \mathcal{V}(\vec{r})=\epsilon_0[{\rm e
    }^{-2\alpha(\vec{r}-l_{ij})}-{\rm e}^{-\alpha(\vec{r}-l_{ij})}],
\end{equation}
where $\epsilon_0$ is the interaction strength, $l_{ij}$ is the mean diameter of particles $i$ and $j$,  and $\alpha$ sets the attraction range (we take $\alpha= 25 l^{-1}$).
The equation of motion is  (\ref{equ:eom}), in which
 the solvent force is $F_{\rm solv} = \sqrt{2\dampbare k_{\rm B}T}\vec{\zeta}$ where $T$ is the temperature, and $\vec{\zeta}$ a unit white noise.
The particles move in a periodic simulation box of dimensions $L_\perp \times L_\perp \times L_{\parallel}$ with $L_\perp=20l$ and $L_\parallel(t=0)=30l$.
The Brownian time is $\tau_{\rm b} = \dampbare l^2/(24k_{\rm B}T)$.
Two natural dimensionless parameters of the model are $\epsilon = \epsilon_0/(k_{\rm B}T)$ which measures the strength of the depletion attraction, and $\damp = \dampbare l/\sqrt{mkT}$ which measures the solvent damping.
In addition, the system undergoes elongation at shear rate $\dot\gamma_0$.  This yields an additional dimensionless parameter $\dot\gamma = \taub \dot\gamma_0$.

In practice, the strain is increased stepwise, with a fixed time period of $\Delta t = 250\tau_{\rm b}$ between steps, so the strain in each step is $\dot\gamma_0\Delta t$.
 When measuring the stress (see below), we average the results over the period $\Delta t$, to reduce the effects of (fast) thermal fluctuations. 
{To confirm that the stepwise elongation does not affect the results, we also performed simulations with continuous elongation, which results in very similar behaviour.}

The numerical simulations are implemented in the LAMMPS package \cite{LAMMPS} in which the natural time scale is $\tau_0=\sqrt{ml^2/k_{\rm B}T}$, this means that $\tau_{\rm b}/\tau_0 =  \lambda/24 \approx 0.42 $ for the parameters used here.  The integration time step is $10^{-4}\tau_0$.
The natural unit of both pressure and stress is $k_{\rm B}T/l^3$; all results for such quantities are quoted relative to this baseline. 

\subsection*{Preparation of the model gel strand}

To set up the arm before elongation, we first initialize a bulk simulation of the model colloid in the NPT ensemble with a low interaction strength $\epsilon=0.01$ and a constant pressure.  {We set $P_0=0.16$
and slowly increase the attractive strength  to  $\epsilon=2.5$, using steps of $\Delta\epsilon=10^{-6}$ every $2 \tau_b$. This causes the volume fraction to increase to $\phi\approx0.59$, the system remains homogeneous because this isobaric transformation not enter the spinodal decomposition regime~\cite{royall2018vitrification}.} We allow this dense homogeneous fluid to relax for a time $\tau_{\rm glass}$.

We then switch to the NVT ensemble and instantaneously adjust  $\epsilon$ to the desired value, in the range $4.5-10$. The strong attractive interactions induce additional dynamical arrest and the system forms a glass-like system. 
After this point, we allowed the simulations to relax for another time $\tau_{\rm glass}$. The system is now inside the liquid-vapour binodal, but the glassy dynamics are slow enough that phase separation is not observed.  (We performed simulations with $\tau_{\rm glass}$ up to $10^5\tau_{\rm b}$, the results depend very weakly on this parameter.   Results are shown for $\tau_{\rm glass} = 10^4\tau_{\rm b}$.)

The result is a homogeneous glassy state with volume fraction $\phi\approx 0.59$.  Our initial strand is obtained by excising a cylinder of radius $r_0$ from this system, after which we run dynamics for approximately $1000\tau_{\rm b}$, to allow the system to relax any features that are artefacts of cutting out the cylinder.  We then start the elongation process.

\subsection*{Stress measurement}

Our measurements of local stress use
a volume-averaged representation of the Irving-Kirkwood  stress~\cite{yang2012generalized,smith2017towards,hardy1982formulas}.  Write $\vec{p}_i$ for the momentum of particle $i$, also $\vec{r}_{ij}$ for the vector connecting particles $i$ and $j$, and $\vec{f}_{ij}$ the corresponding interparticle force.  
Now consider a spatial region $\Omega$ whose volume is $|\Omega|$.  The $\mu\nu$ component of the IK stress for that region is 
\begin{equation}
   \sigma_\Omega^{\mu\nu}= \frac{1}{|\Omega|} \left[ \sum_{i=1}^N\frac{1}{m_i}p_i^\mu p_i^\nu\vartheta_{i,\Omega}+\frac{1}{2}\sum_{i=1}^N\sum_{j\neq i}^N\vec{r}_{ij}^\mu\vec{f}_{ij}^\nu\varphi_{ij,\Omega} \right]\; ,
   \label{equ:IK}
\end{equation}
where $\vartheta_{i,\Omega}=1$ if particle $i$ is in $\Omega$ and zero otherwise; similarly $\varphi_{ij,\Omega}$ is the fraction of the straight line connecting particles $i,j$ that lies within $\Omega$. 

Taking $\Omega$ to be the entire simulation box $\Omega^*$ gives the total stress $\Sigma$, which can also be computed from the virial.  The tensile force is then $F^z = L_\perp^2\sigma_{\Omega^*}^{zz}$.

For a local measurement of stress at point $\vec{r}$, we take $\Omega$ to be a small cube of side {$l_{\rm IK}$}, centred at $\vec r$.  The resulting stress is denoted by $\sigma(\vec{r})$.
For local measurements, we take $l_{\rm IK} = 1.5l$, which is sufficiently small to allow a local measurement, but sufficiently large to avoid numerical uncertainties due to thermal fluctuations (see Appendix).  

As discussed in the main text, it is sometimes convenient to divide the system into $n_{\rm seg}=20$ segments along the $z$-direction, each of which has volume $V_z=L_\perp^2 L_{\parallel}/n_{\rm seg}$.
Then define $N_Z$ as the number of particles in segment 
$Z$.  Taking $\Omega$ in \ref{equ:IK} to be one of these segments gives the tensile force in the arm, divided by the cross-sectional area of the simulation box (which is $L_\perp^2$).  However, the physically-relevant stress is the tensile force divided by the cross-sectional area of the arm. 
{This is obtained by rescaling \eqref{equ:IK}:
\begin{equation}
\overline{\sigma}^{\mu\nu}(Z) = \frac{V_z}{V_s}
\sigma^{\mu\nu}_{\Omega_Z}
\end{equation}
where $\Omega_Z$ is the segment of the box at position $Z$ and $V_s = N_Z \pi l^3/(6\phi)$ is the estimated volume occupied by the arm, within that segment.} (Here $\phi=0.59$ is the volume fraction within the arm, so $\pi l^3/(6\phi)$ is the mean volume per particle there.)

Note that the local IK stress is derived directly from the  equations for momentum conservation.  As such, it accurately reflects the fact that a locally stable gel strand (whose structure is not changing with time), satisfies local force balance $\operatorname{div}\sigma(\vec{r}) = 0$ at every point $\vec{r}$, {that is,
\begin{equation}
\sum_{\mu\in\{x,y,z\}} \frac{\partial}{\partial r^\mu} \sigma^{\mu\nu}(\vec{r}) = 0
\end{equation}
which holds for $\nu=x,y,z$.
See Appendix for further details.}  It is not possible to build a local virial stress with this property.  (The IK method is not the only way to obtain such a stress tensor, but it is a convenient one~\cite{smith2017towards}.)

\subsection*{Stress anisotropy}

The second invariant of the stress tensor measures the anisotropy of the stress, as a scalar quantity that is independent of the orientation of the coordinate system:
     \begin{align}
     J_{2,\Omega}=\frac{1}{2}\text{tr}\Big(\Big[\tens{\sigma}_\Omega-\frac{1}{3}\text{tr}(\tens{\sigma}_\Omega)\Big]^2\Big) \; .
     \end{align}
This quantity is zero if $\sigma$ is proportional to the identity, as would be expected in the bulk of a simple fluid.  In a region $\Omega$ with large anisotropic stresses then $J_2$ will be large.  {The von Mises criterion~\cite{timoshenko1956strength,sica2020mises} for failure of solid materials states that breakage will occur when the local $J_2$ exceeds a threshold.}

For arms under tension with homogeneous stress, the dominant element of $\sigma_\Omega$ is $\sigma^{zz}\approx \sigbarz(Z)$, leading to $J_2 \propto \sigbarz(Z)^2$.  However, in an arm like the one in Fig.~\ref{fig:Macro}(e) with large residual stresses, typical elements of $\sigma_\Omega$ have absolute values larger than $\sigbarz(Z)$, leading to a much larger value of $J_{2,\Omega}$.  It is convenient to average this quantity over a segment of the arm, as
    \begin{align}
\overline{S}(Z)=\frac{l_{\rm IK}^3}{V_S}\sum_{\Omega\in \Omega_Z}\sqrt{J_{2,\Omega}}.
\label{equ:bar-S}
    \end{align}
where the sum runs over cubic regions of size $l_{\rm IK}^3$, within segment $Z$.  Since $J_{2,\Omega}$ is a (non-negative) measure of anisotropy, one sees that $\overline{S}$ captures anisotropic stress fluctuations within the arm.  (Such fluctuations are averaged away in the cross-sectional stress $\sigbarz$.)

\subsection*{Bond-breaking correlation function}

To measure local particle rearrangements, we define
$b_{ij}(\Delta;\gamma)=1$ if particles $ij$ are within a distance $\Delta$ of each other, when the accumulated shear strain is $\gamma$.    Then the fraction of neighbours of particle $i$ that are lost between strains $\gamma$ and $\gamma+\Delta\gamma$ is \cite{scalliet2022thirty}
\begin{equation}
c_i(\gamma,\Delta\gamma)=\frac{\sum_jb_{ij}(\Delta;\gamma)b_{ij}(\Delta;\gamma+\Delta\gamma)}{\sum_jb_{ij}(\Delta;\gamma)} 
\end{equation}
We take $\Delta=1.4l$ throughout as that is the average cut-off range of particle interaction (other values would have given qualitatively the same results).
A correlation function $C_{\rm B}(\gamma,\Delta\gamma)$ is then obtained by averaging $c_i$ over all particles in a suitable region, which we take here to be the segment of the system with position $Z$.

\subsection*{Topological cluster classification}

We analysed particles' local environments using the TCC. 
For each particle $i$, this yields: 
(i) its number of neighbours $n_{\rm n}$; 
(ii) the number of fully-bonded tetrahedra in which it participates $n_{\rm tet}$;
(iii) the numbers of triagonal and pentagonal bipyramids in which it participates, $n_{\rm tb}$ and $n_{\rm pb}$ respectively.  Details and parameters of the TCC are the same as \cite{malins2013tcc}.  These quantities were then averaged over particles in a suitable region (typically the segment of the system with position $Z$).

\subsection*{Acknowledgements} 
We thank Daan Frenkel, Camille Scalliet, Amin Doostmohammadi, Abraham Mauleon-Amieva, Rui Cheng, and Malcolm Faers for helpful discussions.  This work was supported by the EPSRC through grants EP/T031247/1 (KT and RLJ) and 	EP/T031077/1 (CPR and TL). In the later stages of the project, KT also received funding from the European Union's Horizon 2020 research and innovation programme under the Marie Sklodowska Curie grant agreement No. 101029079.



\appendix

\renewcommand{\thefigure}{A\arabic{figure}}
\setcounter{figure}{0}
\renewcommand{\theequation}{A\arabic{equation}}
\setcounter{equation}{0}

\newcommand{\stressDef}{\ref{equ:IK}}
\newcommand{\stressFig}{\ref{fig:Macro}}
\newcommand{\introFig}{\ref{fig:multiple_runs}}

\section*{Appendices}

\begin{figure}[b] 
    \centering
    \includegraphics[width=0.47\textwidth]{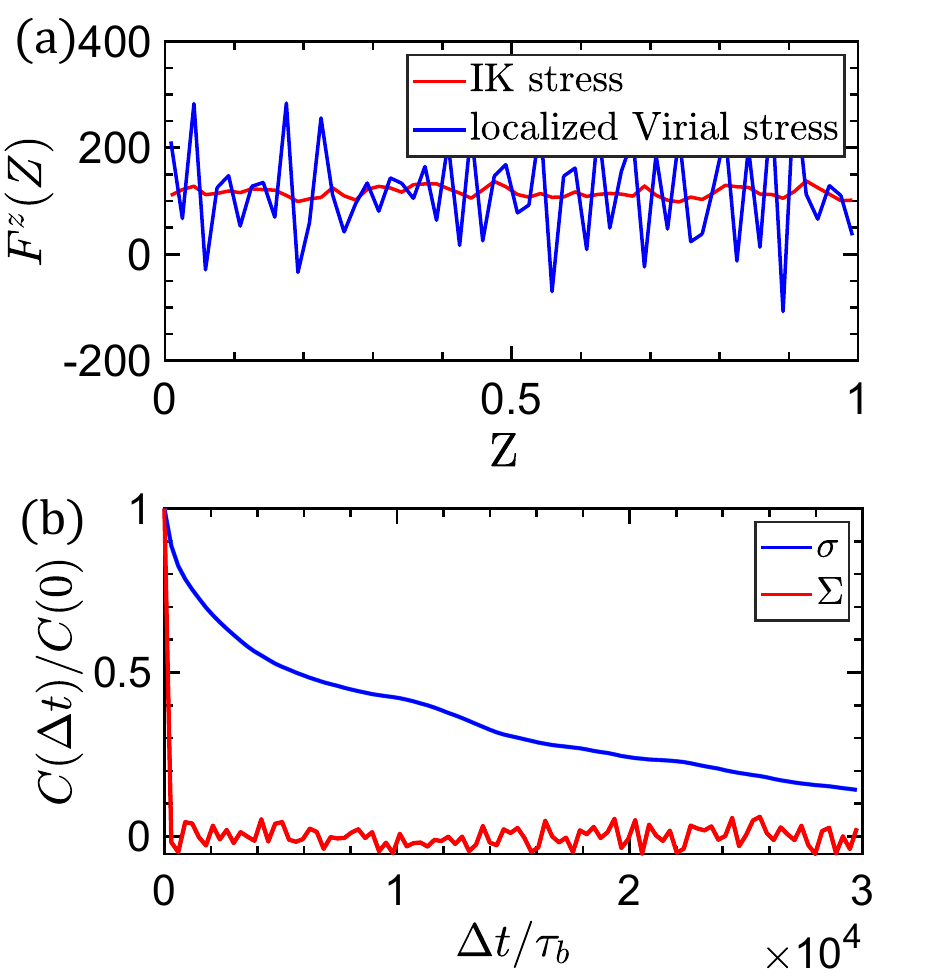}
    \caption{
(a) The tensile force $F(Z)=\int_x\int_y\sigma^{zz} dx dy$ calculated from the Irving-Kirkwood stress and the localized virial stress. Data correspond to a single trajectory of Fig \ref{fig:multiple_runs}({d}) at $\gamma=0.2$, without averaging over the z direction. 
(b) Time-correlation function of the global and local stress. Data used in (b) was obtained from an unstrained strand at the same standard parameters as (a). 
}
\label{fig:method1}
\end{figure} 

\noindent
These Appendices contain additional results and analysis to further justify the methods and conclusions of the main text.

\vspace{6pt}

\noindent
Appendix~\ref{sec:stress-details} discusses our measurements of local stress.  
Appendix~\ref{sec:cryst} discusses numerical simulations of elongation of a crystalline arm, for comparison with the amorphous gel strands considered in main text.
Appendix~\ref{sec:continuum} discusses the continuum rheological model of elongation and necking, including the fitting to numerical data.
\vspace{6pt}


\section{Stress measurements}
\label{sec:stress-details}

\begin{figure*} 
    \centering
    \includegraphics[width=0.9\textwidth]{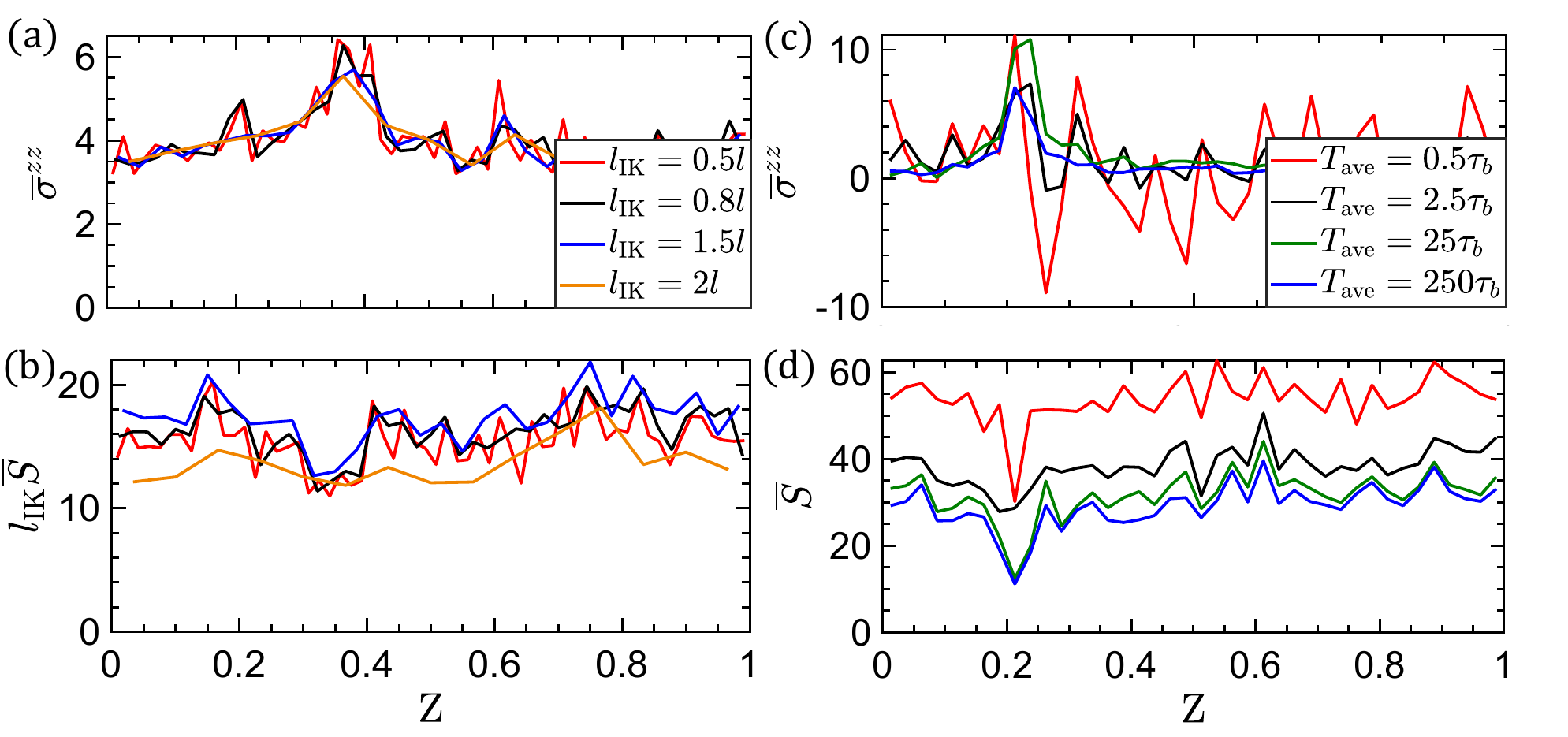}
    \caption{
(a) The average cross-sectional tensile stress $\overline{\sigma}^{zz}$, measured as a volume average over a region of size $L_\parallel^2\times l_{\rm IK}$. Averaging over larger regions reduces the fluctuations, but the increased stress in the neck is clear in all cases (the neck is at $Z\approx 0.35$, in this case).
(b)~Corresponding measurement of anisotropy of the local stress, as a function of the averaging volume (see text for a discussion).  [The coloring of lines is the same as in panel (a).]
(c)~Tensile stress, estimated by time-averaging over a time $T_{\rm ave}$.  Increasing $T_{\rm ave}$ smooths the data, but does not affect the signature of the neck, where $\sigbarz$ increases (at $Z\approx 0.2$ in this case).
(d)~Corresponding stress anisotropy $\overline{S}$: time-averaging reduces the effect of fast anisotropic fluctuations: for the larger averaging times, one sees a stable signal that comes from long-lived residual stresses.
[The coloring of lines is the same as in panel (c).]
Notes: blue lines correspond to the parameters used for all stress measurements outside this Figure; consistent with this, we take $T_{\rm ave}=250\taub$ in (a,b) and $l_{\rm IK}=1.5l$ in (c,d).  Panels (a,b) are taken from a trajectory where the neck appears at $Z\approx 0.35$; panels (c,d) are from a different trajectory where the neck is at $Z\approx 0.2$.  Both trajectories have the same parameters, which are those of Fig.~1(c,d).
}
\label{fig:method2}
\end{figure*}

\subsection{Validation of the Irving-Kirkwood stress}

As discussed in the main text, we use the Irving-Kirkwood (IK) stress $\sigma$ throughout this manuscript to measure the stress on different length scales. Derivations of the IK stress are given in \cite{irving1950statistical,yang2012generalized, dolezal2022mechanical}, and implementation methods are shown in \cite{smith2017towards}. In this manuscript, we use the volume averaged stress from \cite{smith2017towards}.
This approach ensures that the stress is self-averaging when measured over large volumes $\Omega$, which helps to reduce statistical uncertainties.  In particular taking $\Omega$ in (\stressDef) as the full simulation box, the IK stress reduces to the standard virial stress, which is
\begin{equation}
   \Sigma^{\mu\nu}= \frac{1}{V} \left[ \sum_{i=1}^N\frac{1}{m_i}p_i^\mu p_i^\nu+\frac{1}{2}\sum_{i=1}^N\sum_{j(\neq i)}^N{r}_{ij}^\mu{f}_{ij}^\nu \right] \; .
   \label{equ:Vir}
\end{equation}

For local stress measurements, it is possible to define a local virial stress by restricting the sums in (\ref{equ:Vir}) to particles in a particular region.  However, this choice does not ensure that stress gradients cause changes in local momentum, nor that force-balanced systems have $\operatorname{div}\sigma(\vec{r}) = 0$.
To see this, we compute the (local) tensile force  $F^z(Z)$ in the strand by taking $\Omega$ in (\stressDef) as a segment at position $Z$.  Force balance requires that $\partial F^z/\partial Z=0$, up to small corrections due to the thermal fluctuations.  Fig.~\ref{fig:method1}(a) shows that this requirement is obeyed to high accuracy for the IK stress, but it fails for the virial.

The other (non $zz$) components of the stress tensor also have persistent non-zero values (these are local residual stresses).  To illustrate that these components come from the amorphous structure of the arm (instead of fluctuating rapidly on the time scale of velocity fluctuations),  we calculate the time correlation functions of the local and global stress:
\begin{multline}
    C_{\Sigma}(\Delta t)=\sum_{\mu\nu\neq{zz}}\Big[ \big\langle \Sigma^{\mu\nu}(t)\Sigma^{\mu\nu}(t+\Delta t) \big\rangle
    \\
     - \big\langle \Sigma^{\mu\nu}(t) \big\rangle
       \big\langle \Sigma^{\mu\nu}(t+\Delta t)\big\rangle
       \Big]
\end{multline}
and
\begin{multline}
C_\sigma(\Delta t)=\frac{1}{V}\int_V\sum_{\mu\nu\neq{zz}}
\Big[ 
\big\langle\sigma^{\mu\nu}(\vec{r};t)\sigma^{\mu\nu}(\vec{r};t+\Delta t) \big\rangle
\\
- \big\langle \sigma^{\mu\nu}(\vec{r},t)\big\rangle
       \big\langle \sigma^{\mu\nu}(\vec{r},t+\Delta t)\big\rangle
\Big] 
d\vec{r}.
\end{multline}
{In these sums, both $\mu$ and $\nu$ run over the three Cartesian directions $x,y,z$, but the term $\mu=\nu=z$ is excluded.}

Figure \ref{fig:method1}(b) compares the global and local stress correlation functions, during simulations of an arm without any elongation.  The local stress correlations decay slowly, showing that the residual stresses are long-lived.  We attribute this to slow structural changes in the (non-equilibrium) strand,  thousands of Brownian times are required for these changes to become significant.  By contrast, all elements of the global stress are very small (except $zz$).  {As a result, the dominant contributions to $C_\Sigma$ are fast thermal fluctuations, so this correlation function decays quickly.}

It is well-known that microscopic expressions for the local stress tensor are not unique~\cite{smith2017towards}.  We briefly discuss three aspects of this issue.
First, since these systems have significant residual stresses, the value of $\sigma$ depends on the scale at which it is measured.  Our local stress tensor is measured by taking $\Omega$ in (\stressDef) to be a cubic box of side $3l/2$.  This length scale is chosen for numerical convenience: taking larger boxes tends to smooth out the stress, but smaller boxes lead to a noisy signal (see Sec.~\ref{sec:ave-stress} for further discussion of this point).
Second, the terms arising from pairwise forces in the IK stress are evaluated by considering a linear interaction path between the particles.  Other paths are possible but the linear path is a simple and convenient choice.  For long-ranged forces, the choice of path can significantly affect the stress, but for these short-ranged Morse potentials, such effects are small (as long as a reasonable path is used).
Third, one might (in principle) also shift all stress values by a global constant (the reference pressure), here we insist that a very dilute colloidal suspension has a vanishingly small (osmotic) pressure, so this constant is zero.

\subsection{Averaging the stress}
\label{sec:ave-stress}

Throughout this work, we report IK stresses that are averaged over a time period $T=250 \tau_{\rm b}$.  We recall from Fig.~\ref{fig:method1}b that this time scale is small enough that the stress does not relax significantly, but we do find that the averaging process reduces statistical noise.
(Specifically, we compute the stress at time points separated by $\tau_{\rm b}/2$ and we average over 500 such time points to obtain the reported values.)  
The justification of this averaging relies on of the slow time evolution of the local stress (recall Fig.~\ref{fig:method1}), which is in turn due to the dynamically-arrested (solid) structure of the arm.  In the bulk of a simple fluid, this kind of averaging would yield instead an isotropic stress tensor, although the situation would be more complicated in systems with interfaces or applied forces~\cite{braga2018pressure}. 
As noted above, we also measure a volume-averaged local stress, based on a cubic box of size $l_{\rm IK}=1.5l$.  

Fig.~\ref{fig:method2} illustrates the effects of these averaging procedures, for representative arms.  For the stress itself, larger averaging times and volumes lead to smoother signals, as expected.  This is shown in Fig.~\ref{fig:method2}(a,c).
We also computed the stress anisotropy $\overline{S}(Z)$, as defined in \eqref{equ:bar-S}.  In general, larger averaging times and volumes suppress the effects of anisotropic fluctuations.  To faithfully capture these fluctuations (as in Fig.~\ref{fig:Macro}), we use intermediate length and time scales for averaging, as we now discuss.

We first consider effects of volume-averaging, over cubes of volume $l_{\rm IK}^3$.  The IK stress distributes the contribution of each pair of particles along a line connecting them.  For very small volumes, this results in many cubes with no contribution to the stress, and others with large anisotropic contributions.  The resulting mean stress is independent of $l_{\rm IK}$ but the anisotropic fluctuations behave as $\overline{S}\sim l_{\rm IK}^{-1}$ for small $l_{\rm IK}$.  Fig.~\ref{fig:method2}(b) plots $l_{\rm IK} \overline{S}(Z)$ for various sizes $l_{\rm IK}$.  Increasing the box size up to $l_{\rm IK} \approx 1.5l$ suppresses a noisy contribution from thermal fluctuations, helping to reveal the reduction in $\overline{S}$ near the neck.  For larger averaging volumes, the anisotropic fluctuations of the stress are significantly reduced: this is because the anisotropic residual stresses illustrated in Figs.~\ref{fig:Macro}(e,f) are being averaged away.  (Indeed, averaging over the entire cross section will eventually yield the picture of Figs.~\ref{fig:Macro}(a,b), where anisotropic fluctuations are much smaller.)  Hence we choose $l_{\rm IK}=1.5l$ for our measurements, which is large enough to suppress fluctuations from thermal noise, without averaging away the physically-relevant residual stresses.

For time averaging, the picture is simpler.  Taking an average over $T_{\rm ave}=250\taub$ effectively suppresses fast thermal fluctuations in the measured stress [Fig.~\ref{fig:method2}(c)].  Since these fluctuations are anisotropic, the time-averaging also suppresses the anisotropy [Fig.~\ref{fig:method2}(d)].  There is a broad range of times around $T_{\rm ave}=250\taub$ where these signals are stable.  For much larger times, one loses resolution in time due to local stress relaxation (from Fig.~\ref{fig:method1}, this happens on time scales are $\gtrsim 1000\taub$).  Hence we choose $T_{\rm ave}=250\taub$ for our measurements, as a sensible compromise between time-resolution and noise reduction.

\begin{figure} 
    \centering
    \includegraphics[width=0.45\textwidth]{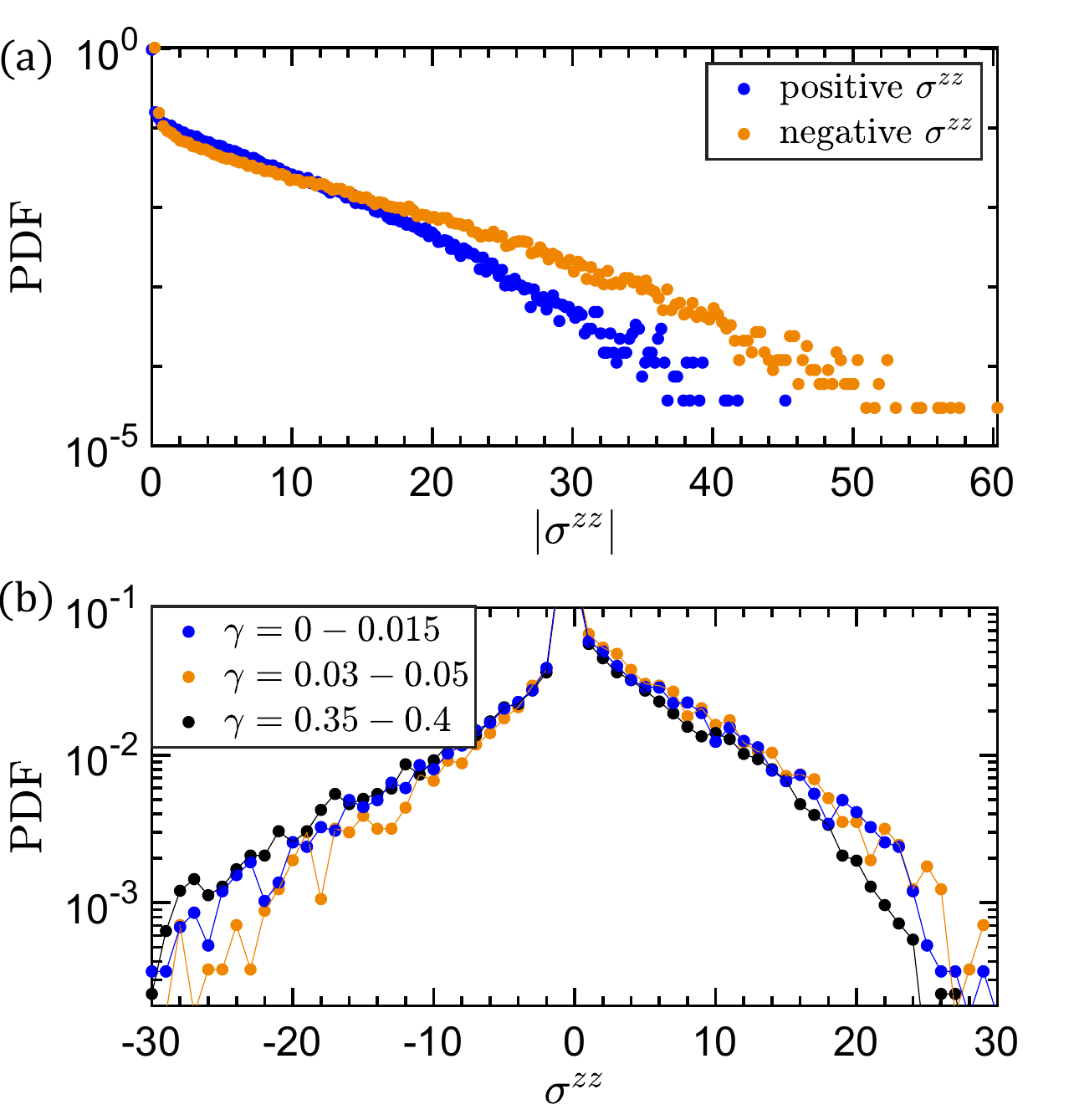}
    \caption{Distributions of the local stress $\sigma^{zz}$.
(a) Comparison of the distributions for positive and negative stress values, before any elongation has occurred. Data is found by measuring for $10^6 \tau_b$ without elongation.
(b) Stress distribution evolution far away from the neck ($Z=1$) near the start of the simulation ($\gamma=0-0.05$), just after the system has reached its stress plateau ($\gamma=0.03-0.05$), and just before necking occurs ($\gamma=0.35-0.4$). Data has been averaged over 6 trajectories from Figure \ref{fig:multiple_runs} in the main text.
}
\label{fig:method3}
\end{figure} 

\subsection{Additional information about stress distribution}

Fig.~\stressFig(h) of the main text shows the distribution of the local stress (more precisely, its $zz$ component).  We emphasized the differences in this distribution between the neck region and the bulk of the strand.  We present here some additional information on these distributions.

At initialization, long-lived residual stresses are quenched into the system: the tails of these distributions are roughly exponential, with different decay rates for positive and negative stress. Fig.~\ref{fig:method3}(a) shows that positive stresses are more common, but the negative stress distribution has the fatter tail.  Overall, the total stress (which coincides with the average of this distribution) is lightly positive at initialization, corresponding to a tensile stress.

Fig.~\ref{fig:method3}(b) shows the stress in the bulk of the arm (away from the neck), as the arm is stretched.  The tensile force increases during elongation corresponding to an increase in the average of this distribution.  However, this shift is weak, in comparison with the residual stresses that are already present, so it has a weak effect on the stress distribution.

\begin{figure} 
    \centering
    \includegraphics[width=0.49\textwidth]{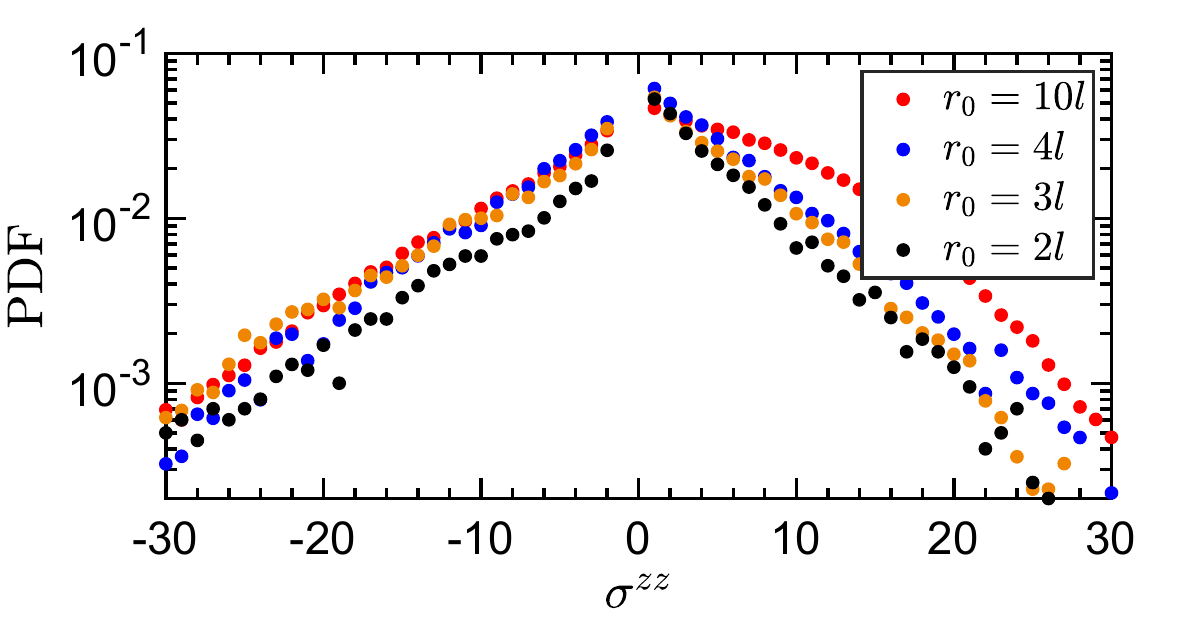}
    \caption{
 The stress distribution for different starting radii with no deformation.  Data is found by measuring for $10^6 \tau_b$ without straining. Other parameter values are those of Fig.~\ref{fig:multiple_runs} of the main text.
}
\label{fig:method4}
\end{figure}

We also note that the change in stress distribution for the neck is not just an effect of its reduced thickness (which leads to a change in the ratio of surface to bulk).  We finitialized strands with different radii: they all have similar non-Gaussian distributions of the stress (Figure \ref{fig:method4}).

\section{Elongation of a crystalline arm}
\label{sec:cryst}

\begin{figure} 
    \centering
    \includegraphics[width=0.45\textwidth]{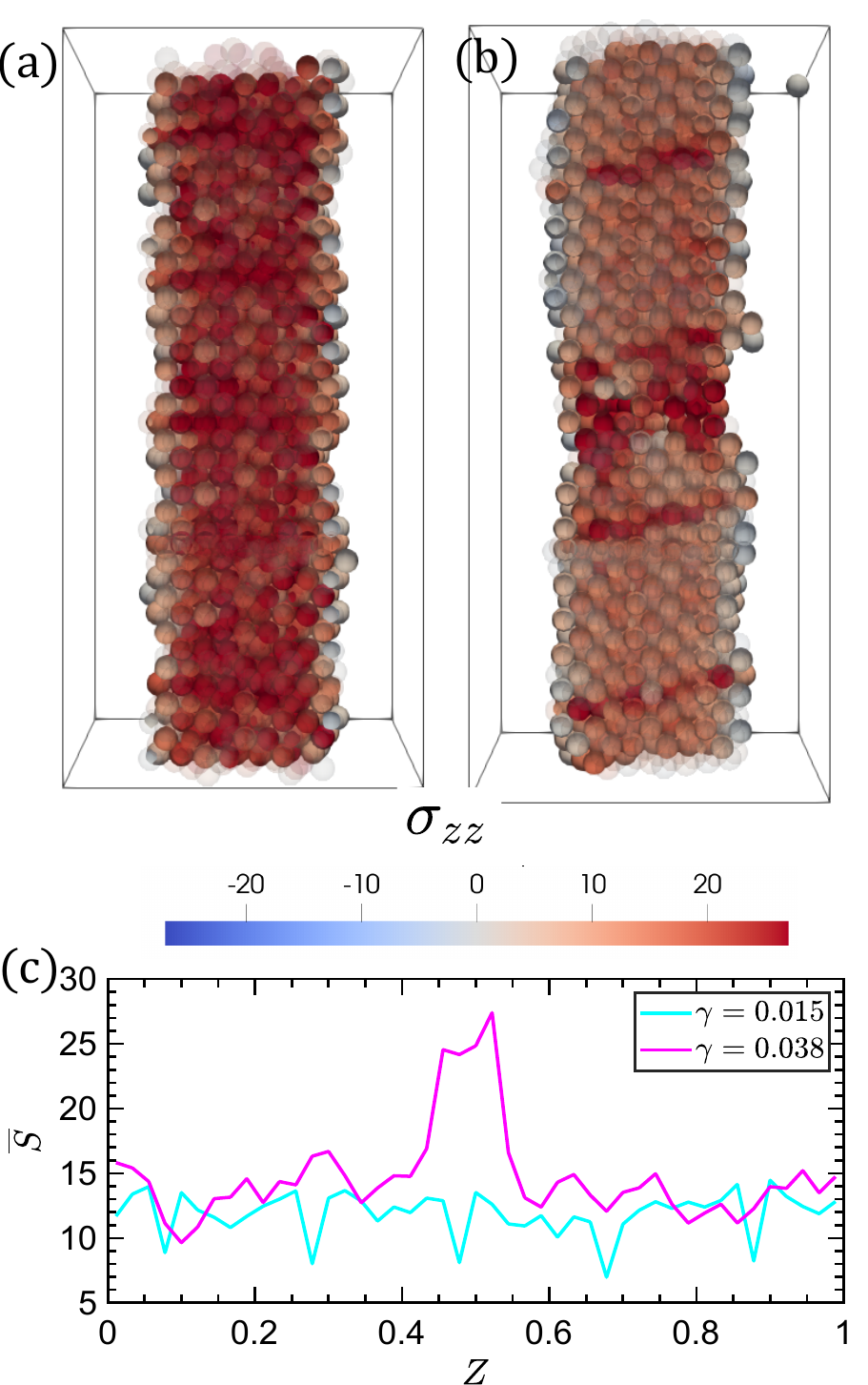}
    \caption{
The local stress $\sigma^{zz}$ (a) before neck formation $(\gamma=0.015)$ and (b) after neck formation $(\gamma=0.038)$ for a system where particles are initialized on an FCC lattice.
(c) The corresponding measurement of anisotropy in the  stress tensor $\overline{S}$. Other parameter values are those of Fig.~\ref{fig:multiple_runs} of the main text.
}
\label{fig:Crystal}
\end{figure}

As a point of comparison for the strand elongation discussed in the main text, we also performed similar experiments on a crystalline arm.  
We initialised a face-centred cubic (FCC) crystal from which we excised a cylindrical strand.  On elongation, a few  plasticity events were observed, after which the strand broke by a sudden mechanism resembling brittle failure.  Results are shown in Fig.~\ref{fig:Crystal}.
Looking at the mesoscopic stress, we can see that the magnitude of the local stress $\sigma_{zz}$ increases at the failure point (Figure \ref{fig:Crystal}(a)), as expected because the arm is thinnest there.  

However, a striking difference between the crystalline and amorphous arms  is the absence of residual stresses in the crystalline case.  
As a result, the stress is relatively homogeneous.  On computing the stress anisotropy in the crystal, we find that $\overline S(Z)$ is \emph{largest} in the neck.  This stands in contrast to amorphous arm, where the stress was more homogeneous in the neck, and $\overline S(Z)$ was smaller there.

\section{Continuum model}
\label{sec:continuum}

The theory presented in the manuscript is based on the analysis of \cite{moriel2018necking}, although it can also be interpreted in the framework of \cite{hoyle2015age,hoyle2016criteria}. 

\subsection{Definition}

The model uses a thin-filament approximation so the strand is modelled in one-dimension (oriented along the $z$ direction). 
Define a local velocity $\partial_zV(z,t)=\dot{\gamma_L}$ with $\dot{\gamma_L}(z,t)$ the local strain rate, which differs in general from the externally imposed strain rate $\dot\gamma$. 
We also assume that only the area of the strand $A(z,t)$ is important, when considering mass conservation. We will write our equations in the co-extending frame of imposed strainrate $\dot{\gamma}$.  Hence we define~\cite{hoyle2016criteria}
\begin{align}
    Z&=z \exp(-{\dot{\gamma}}t),\\
    v(Z,t)&=V(z,t)\exp(-{\dot{\gamma}}t),\\
    a(Z,t)&=A(z,t)\exp({\dot{\gamma}}t),
\end{align}
where we have normalized $z$ by the strand length $L_\perp(t)$, to keep the new spatial variable $Z$ between 0 and 1. 
Transformation to this co-extending frame, the mass conservation equation equation becomes 
\begin{equation}
\frac{\partial a}{\partial t}=-(\dot{\gamma}_L-\dot{\gamma})a.
\label{equ:mass}
\end{equation}  
(Here and throughout we neglect advective terms from the change of frame, which is valid for slow elongation rates.)

To account for forces in the arm, we write $W(Z,t)$ for the tensile stress in the $z$ direction, averaged over the cross-section of the strand.  The tensile force is then $aW$ and the system remains force-balanced at all times, so 
\begin{equation}
\frac{\partial}{\partial Z}(a W)=0 .
\label{equ:forceBal}
\end{equation}
We assume that the stress evolves with an elastic loading term and relaxes due to local plasticity events which yields
\begin{align}
\frac{\partial W}{\partial t}= G\left[ \dot{\gamma}-p(W,\chi)\right], \label{eq:W_stress-SM}
\end{align}
where $G$ is the elastic modulus and $p$ the rate of plastic relaxation, which depends in turn on an internal plasticity field $\chi$ (see below).
More specifically, we follow~\cite{moriel2018necking}: $p$ is zero when $W$ is less than a yield stress $\sigY$, and $\chi$ behaves like a temperature, which controls the rate of ``activated'' plastic events.  {(As discussed in the main text, such relationships are familiar from shear transformation zone theory~\cite{falk1998}.)}
Hence, 
\begin{equation}
p(W,\chi)=
\frac{W-\sigY}{\tau_p(\chi)\sigY} \theta \left(W-\sigma_y\right)
\end{equation}
where $\theta$ is the Heaviside (step) function and $\tau_p(\chi)=\tau_{\rm ref}\exp{(1/\chi)}$, where $\tau_{\rm ref}$ is a reference time scale.

In the simplest case we take $\chi$ as a constant parameter, but this is not sufficient to capture the stress-strain relationship observed in simulations.  Instead we again follow~\cite{moriel2018necking} in promoting $\chi$ to a dynamical variable whose relaxation is controlled by the plastic time scale $\tau_p$.  Specifically, we take
\begin{align}
\frac{\partial\chi}{\partial t}=\frac{W}{\sigY}p(W,\chi) [\chi_\infty-\chi].
\label{eq:chi-SM}
\end{align}
where the parameter $\chi_\infty$ sets the steady-state value of $\chi$.  

\subsection{Homogeneous solution: qualitative behaviour and fitting to numerical data}

We first consider homogeneous solutions of this model, that is, solutions where strain, area, $W$ and $\chi$ are independent of $Z$.  
At early times,  we  have ${W}(t)=W_{t=0}+G\dot{\gamma}t$, which holds for times short enough that $W<\sigY$.   For long times, we have
 $W(t) \to\sigY/(1-\dot{\gamma}\tau_p)$ where $\tau_p$ is the plastic time in the steady state.

As discussed in the main text, an important question is: when $W$ reaches the yield stress $\sigY$ and the plastic activity starts to occur, does $\chi$ relax quickly to $\chi_\infty$, or is the time scale for this relaxation compete with the relaxation of $W$ to its steady-state value? This determines whether an stress overshoot is observed.

The resulting model has several parameters, which we fit to the data of Fig.~\introFig(e) in a multi-step procedure.  We identify $W$ with $\sigbarz$ (averaged over $Z$) and we fit the data at early times to $\sigbarz=W_{t=0}+G\dot{\gamma}t$, which yields $G=420$ and $W_{t=0}=0.5$ [recall that the units of both these quantities are $k_{\rm B}T/l^3$].  We then consider the plateau of $\sigbarz$ (before significant necking occurs): we fit the plateau height to $W=\sigY/(1-\dot{\gamma}\tau_p)$ which yields $\tau_p=180\taub$ and $\sigY=3.3$.  Note however that this $\tau_p$ is the steady-state value of the plastic time scale: since $\tau_p=\tau_{\rm ref}{e}^{1/\chi}$ then this is not itself a parameter of the model.  To obtain values for $\tau_{\rm ref}$ and $\chi_\infty$ we fit the stress overshoot that occurs in Fig.~\introFig(e) for  $\dot\gamma=5\times 10^{-4}$: this yields $\tau_{\rm ref}=180\taub$, and $\chi_\infty=80$, as well as the initial condition $\chi_{t=0}=0.65$.  At this point all parameters have been determined for the fitting in Fig.~\introFig(e).

\subsection{Necking as a linear stability of the homogeneous solution}

To analyse necking, we consider the linear stability of the homogeneous solution for $(\dot\gamma_L,a,W,\chi)$.  
We consider a perturbation to the homgeneous solution at wavevector $q$, that is
\begin{equation}
\begin{pmatrix}
\dot{\gamma_L}(Z,t) \\ a(Z,t)  \\ W(Z,t)\\ \chi(Z,t)
\end{pmatrix} 
= 
 \begin{pmatrix}
\dot{\gamma} \\ a_0(t)  \\ W_0(t)\\ \chi_0(t)
\end{pmatrix} +
\begin{pmatrix}
\delta \dot{\gamma}(t) \\ \delta a(t)  \\ \delta W(t) \\ \delta \chi(t)
\end{pmatrix}
\exp(iqZ)
\end{equation}
{where} the $0$-subscripts indicate the solution to the homogeneous equation.
Following~\cite{moriel2018necking}, Eq.~\ref{equ:mass} fixes $\dot\gamma_L$ in terms of $a$ and its time derivative. Linearising Eq.~(\ref{equ:forceBal}) determines $\delta a$ in terms of $\delta W,a_0,W_0$.   Linearising  (\ref{eq:W_stress-SM},\ref{eq:chi-SM}) and using these two relations yields a closed equation for the perturbation as
\begin{equation}
\frac{\partial}{\partial t} \begin{pmatrix}
 \delta W \\ \delta \chi
\end{pmatrix}
 = M \begin{pmatrix}
 \delta W \\ \delta \chi
\end{pmatrix}
\end{equation}
with stability matrix
\begin{align}
M=\begin{bmatrix}
W_0\partial_Wp(W_0,\chi_0)-p(W_0,\chi_0) &  -a_0\partial_\chi p(W_0,\chi_0) \\ 
-\frac{W_0}{a_0}\partial_W \dot{\chi}(W_0,\chi_0) & \partial_\chi \dot{\chi}(W_0,\chi_0)
\end{bmatrix}.
\end{align}
where 
 the notation $\dot\chi$ is a shorthand for the right-hand side of \eqref{eq:chi-SM}, interpreted as a function of $(W_0,\chi_0)$.

For $W_0<\sigY$, there is no plastic flow and $M=0$: the strand responds elastically and preserves its shape under elongation.  
However, for $W_0>\sigY$, the matrix always has one positive and one negative eigenvalue, indicating that the homogeneous solution is linearly unstable and a neck will form.  This situation -- of a single real positive eigenvalue -- corresponds to the slow (or gradual) instability of \cite{ hoyle2015age,hoyle2016criteria}, which is consistent with our numerical simulations.

\bibliography{Bibliography}

\end{document}